\documentclass[pdftex]{jps-cp}
\usepackage{txfonts} 

\title{A Study to Suppress a Sneaking Cosmic Muon Background in the COMET Experiment}

\author{Manabu \textsc{Moritsu}$^{1}$, Hai-Bo \textsc{Li}$^{2}$, Tianyu \textsc{Xing}$^{2}$, Ye \textsc{Yuan}$^{2}$, and Yao \textsc{Zhang}$^{2}$}

\inst{$^{1}$Department of Physics, Kyushu University, Fukuoka 819-0395, Japan \\
$^{2}$Institute of High Energy Physics (IHEP), Chinese Academy of Sciences (CAS), Beijing 100049, China}

\email{moritsu.manabu@phys.kyushu-u.ac.jp}

\recdate{April 8, 2025}

\abst{ 
The COMET experiment, conducted at J-PARC, aims to search for muon-to-electron conversion with an unprecedentedly high sensitivity. 
One of the severest backgrounds in the Phase-I experiment originates from cosmic-ray muons. 
A cosmic muon sneaks into a detector solenoid magnet from a loophole, scattered and leaving a track in a cylindrical drift chamber. 
Among them, a positive muon track with reverse direction may mimic a signal electron of 105 MeV/$c$. 
In order to suppress the sneaking cosmic positive muon background, we developed a method to discriminate the track direction by using track-fitting quality. 
We demonstrated that the positive muon background can be reduced by an order of magnitude. 
In this paper, we will report the methodology, a Monte Carlo simulation and results with prospects. 
} 

\kword{muon, cosmic ray, tracking, drift chamber, J-PARC}

\begin{document}
\maketitle

\section{Introduction}

We are searching for neutrinoless muon-to-electron conversion in a muonic atom to pursue new physics beyond the Standard Model in particle physics. 
The COMET experiment will be conducted at Japan Proton Accelerator Research Complex (J-PARC), aiming at the world's highest sensitivity for the muon-to-electron conversion process \cite{COMET20}. 
The construction of the dedicated beam line, muon source and detectors are in progress \cite{Moritsu22, Moritsu21}. 
The first beam commissioning, called Phase-$\alpha$, has been performed in 2023, delivering an 8-GeV proton beam from J-PARC Main Ring to the COMET experimental facility. 
We have confirmed that low-energy negative muons generated in the backward direction were transported through the superconducting solenoid and stopped at a muon stopping target \cite{Oishi25}. 

In Phase-I of the COMET experiment, we plan to achieve the single event sensitivity of $3 \times 10^{-15}$, which is two orders of magnitude better than the current experimental upper limit \cite{Bertl06}. 
A cylindrical detector system is used to measure a signal electron from muon-to-electron conversion, the momentum of which is monochromatic: 105 MeV/$c$ in the case of aluminum stopping target. 
The detector system consists of a cylindrical drift chamber (CDC) \cite{Sato24} and cylindrical trigger hodoscopes (CTH) in a solenoidal magnetic field of 1~T. 
The detector solenoid (DS) magnet is surrounded by a cosmic-ray veto (CRV) system composed of 4 layers of plastic scintillator \cite{Artikov24} or resistive plate chamber arrays. 

\begin{figure}[t] 
  \centering 
  \includegraphics[width=0.55\linewidth]{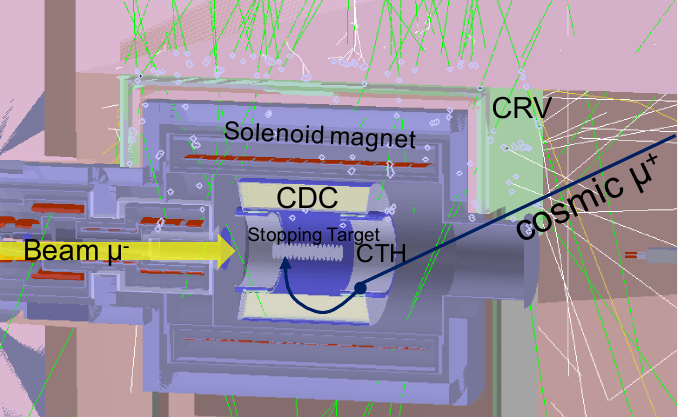}
  \caption{Cross-sectional view of the detector region in COMET Phase-I. An example of the sneaking cosmic muon trajectory from the backward side is also drawn.}
  \label{fig:sneaking_crv1}
  \vspace{5mm}
  \includegraphics[width=0.6\linewidth]{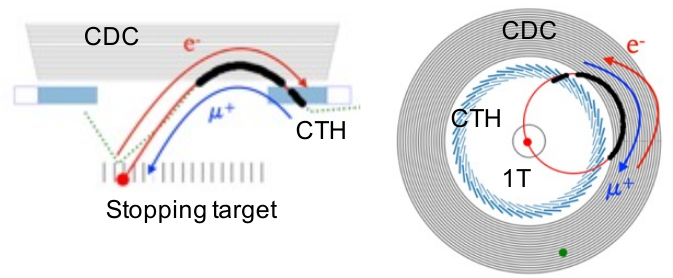}
  \caption{Illustrations of a particle trajectory and detector hits in the longitudinal side view (left) and the transverse view from downstream (right). The red arrows indicate a $\mu$-$e$ conversion signal track with the normal direction; whereas the blue arrows indicate a sneaking cosmic $\mu^+$ track with the reverse direction.}
  \label{fig:sneaking_crv2}
\end{figure}

There are several types of background processes in muon-to-electron conversion such as the decay-in-orbit of a muon in the atomic orbit, other particles contaminating the muon beam, decay-in-flight of high-energy muons and so forth. 
In addition to these, a cosmic-ray background is considered as one of the severest background. 
Cosmic rays, mostly muons, falling from outside of the DS magnet react with the stopping target or vicinity, and knock out electrons. 
If this electron happens to be 105 MeV/$c$, it cannot be distinguished from the signal event. 
To guarantee that it is not caused by cosmic rays, the CRV system is required to have no hits coinciding with the signal timing. 
Whereas the roof and side walls of the DS magnet are designed to be fully covered by CRV, there are loopholes that CRV cannot practically cover due to beam ducts, cable wiring, {\it etc.} as shown in Fig.~\ref{fig:sneaking_crv1}. 
Most of cosmic rays come from above according to the zenith angle, $\theta$, distribution of $\cos^2{\theta}$, but many also enter from the backward side at large angles. 

The cosmic-ray background rate was estimated with a backward Monte Carlo simulation \cite{Niess22} to be $2.4 \pm 0.9$ events for 150 days of the Phase-I operation. 
A typical example of the dangerous events is also illustrated in Fig.~\ref{fig:sneaking_crv1}; 
in this case, the most problematic particle is not a knockout electron but a cosmic-ray muon itself; 
the cosmic-ray muon does not hit CRV, sneaking from the downstream loophole into DS, scattered by the CTH support structure, making hits in CTH and then CDC, and finally reaching the stopping target. 
It should be noticed that this problematic cosmic ray is not a negative muon but a positive muon, because the trajectory of a reversely-going positive charged particle looks almost the same as the signal electron track from a viewpoint of the detector data as illustrated in Fig.~\ref{fig:sneaking_crv2}. 
We call this type of dangerous events as a {\it sneaking cosmic muon ($\mu^+$) background}. 
Hereafter, we define the track direction of the signal electrons as {\it normal} and that of the sneaking cosmic-ray $\mu^+$ as {\it reverse}. 

In this Article, we will discuss methodology to suppress the sneaking cosmic muon background from the viewpoint of detector data analysis, in particular, focusing on a tracking analysis. 
We have developed a track reconstruction program for CDC, and performed a simple Monte Carlo simulation as a proof of the principle.

\section{Methodology}

\subsection{Discrimination of the sneaking cosmic muon background}

Features of the sneaking cosmic muon background is summarized in Table~\ref{tab:feature}, compared with the muon-to-electron conversion signal. 
The problematic background is caused by cosmic positive muons that happen to be 105 MeV/$c$. 
At this momentum, the velocity $\beta$ of muons is 0.71 whereas the electron velocity is almost 1 in a unit of the speed of light $c$. 
As defined above, the {\it normal} track direction represents a counterclockwise rotation viewed from downstream, generating from the stopping target, passing through CDC and reaching CTH; in contrast, the {\it reverse} direction is a clockwise rotation, generating from CTH, passing through CDC and reaching the stopping target (Fig.~\ref{fig:sneaking_crv2}). 

In this subsection, we will discuss a few possible methods to discriminate the signal from the sneaking cosmic background as follows. 

\begin{table}[h] 
    \centering
    \begin{tabular}{lcc}
        \hline
         & $\mu$-$e$ conversion signal & Sneaking cosmic background \\
        \hline
        Particle & $e^-$  & $\mu^+$ \\
        Velocity $\beta$ for 105 MeV/$c$  & 1  &  0.71  \\
        Track direction & Normal  & Reverse  \\
        \hline
    \end{tabular}
    \caption{Features of the $\mu$-$e$ conversion signal and the sneaking cosmic muon background. See the text for the definition of {\it normal} and {\it reverse} directions.}
    \label{tab:feature}
\end{table} 

\subsubsection{Methods using CTH}

In the current detector design, CTH consists of 256 segmented plastic scintillators read by optical fibers with silicon photomultipliers. 
It can be easily thought that if the scintillators are replaced with radiators with the refractive index smaller than 1.4, the Cherenkov radiation can be available for the discrimination. 
There are such materials as water, aerogels {\it etc.}, but, since the detector design has been almost fixed, the implementation would be practically difficult due to the complexity of the installation, trigger efficiency, cost expense and so forth. 

It is capable to utilize the energy deposit in the plastic scintillators of CTH. 
The energy loss of muons in the plastic is slightly larger than that of electrons at 105 MeV/$c$. 
This study was reported in \cite{Fujii24}, and the reduction factor greater than 10 with the signal retention higher than 90\% were demonstrated with a beam test.

\subsubsection{Methods using CDC}

The CDC design was optimized to measure the 105-MeV/$c$ electrons at the 1-T magnetic field with momentum resolution better than 200 keV/$c$ \cite{Sato24}. 
The chamber is operated with a low-$Z$ gas mixture of He (90\%) and iC$_{4}$H$_{10}$ (10\%) to minimize multiple scattering effects. 
CDC consists of 4986 channels of drift cells structured in 20 concentric layers, all of which have stereo angles of alternately $+4$ and $-4$ degrees to effectively estimate the longitudinal position. 
The 105-MeV/$c$ signal electrons typically hit 50--60 cells of CDC. 
The average spatial resolution is expected to be 150~$\mu$m based on a prototype chamber test \cite{Wu21}. 
The construction of CDC was completed, and the performance tests were carried out \cite{Moritsu19, Moritsu20}. 

To discriminate the signal electrons from the sneaking cosmic muons, it seems to be hardly capable to utilize the difference of the energy deposits in CDC. 
The energy loss of muons in a helium gas is accidentally almost the same as that of electrons at 105 MeV/$c$. 
The difference from the CTH case above is caused by the density effect \cite{PDG24}; 
the relativistic rise of the energy loss for electrons is less suppressed in low-density materials, 
and therefore the electrons lose more energy in a helium gas than in a plastic. 

In the end, let us pay attention to the difference of the track directions. 
As described above, compared with the normal track direction of the signal electrons, the sneaking cosmic $\mu^+$ goes with the reverse direction. 
CDC can precisely measure the particle hit positions by using the drift time of ionized electrons generated along the trajectory; 
at that time, CTH provides the time origin to the drift time. 
Note that, to precisely determine the time origin, one should correct the time lag with the time-of-flight between CTH and each CDC hits. 
The track is normally reconstructed assuming that the particle is generated from the stopping target, passes through CDC, and hits CTH. 
In contrast, the sneaking cosmic $\mu^+$ hits CTH first and then passes through CDC. 
This causes an error of up to 10~ns at longest in the time origin of the drift time. 
This miscorrection corresponds to the drift distance of about 250~$\mu$m, using typical drift velocity of 25~$\mu$m/ns in the He:iC$_{4}$H$_{10}$ (90:10) gas mixture. 
The hit position error of 250~$\mu$m is comparable with the CDC spatial resolution of 150~$\mu$m. 
Based on this idea, the quality of the track reconstruction, {\it e.g.} $\chi^2$, should differ depending on which direction the particle track is assumed to be. 

In addition, the following can also be considered. 
A particle loses the energy, typically 1.5~keV per cell, along the CDC trajectory, and the curvature gradually becomes slightly smaller. 
This effect is taken into account by the Kalman filter in the track fitting analysis; 
however, if the assumed track direction is opposite, the correction is made in the wrong direction. 
This also worsens the quality of the track reconstruction.

\subsection{Development of the track reconstruction program}
\label{sec:track_recon}

In order to demonstrate the principle of the track-direction identification, we developed a track reconstruction program. 
The particle tracks around 100~MeV/$c$ are curled in CDC, which adds more complication to the tracking compared with straight tracks. 
In actual experimental analysis, to reduce noise hits, the CDC data are firstly processed into the hit classification using the boosted decision tree and track finding using the Hough transformation or deep learning \cite{Kaneko24}; 
After that the signal hits are processed into track fitting. 
Although one third of the tracks are actually multi-turn tracks \cite{Zhang19}, we concentrated on only single-turn tracks in this paper. 

We used a track fitting algorithm implemented in the offline software framework for the COMET experiment. 
The track fitting algorithm is based on a generic track fitting toolkit, GENFIT2 \cite{Hoppner10, Bilka19}, providing modular packages independent of experiments, known to be suitable for relatively low-energy experiments. 
We adopted, as a fitter option, Deterministic Annealing Filter, which allows competition between measurements and rejection of fake measurements.

\section{Simulation and Results}

\begin{figure}[tbh] 
  \centering 
  \includegraphics[width=0.7\linewidth]{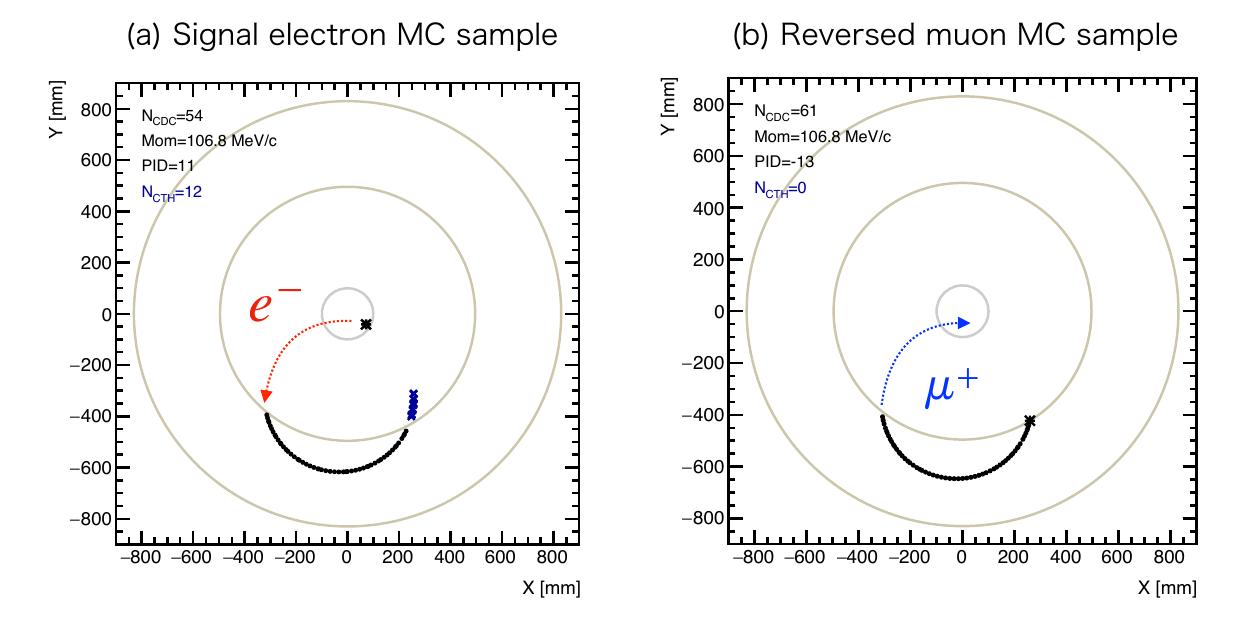}
  \caption{Typical event displays in the Monte Carlo simulation for (a) the signal electron samples and (b) reversed positive muon samples, viewed from the downstream.}
  \label{fig:event-display}
\end{figure}

We generated Monte Carlo simulation samples for both the signal electrons and the sneaking cosmic muons. 
It should be noted that we did not generate noise hits in this simulation. 
For the signal electron samples, electrons were isotropically generated from the stopping target with a uniform momentum distribution of 100--110 MeV/$c$. 
The tracks were required to reach at least the fifth layer of CDC and to make the CTH hits. 
Only the single-turn tracks were used in the subsequent processes. 
The sneaking cosmic muon samples were simulated by using a backward Monte Carlo method, where positive muons were generated from the CTH hit points of the signal electron samples with the reversed momenta. 
The tracks were required to reach at least the fifth layer of CDC and to hit the stopping target. 
Typical event displays for the electron and reversed muon samples are shown in Fig.~\ref{fig:event-display}. 
The CDC hit positions were smeared with the expected spatial resolution of 150~$\mu$m. 

\begin{figure}[tbh] 
  \includegraphics[width=\linewidth]{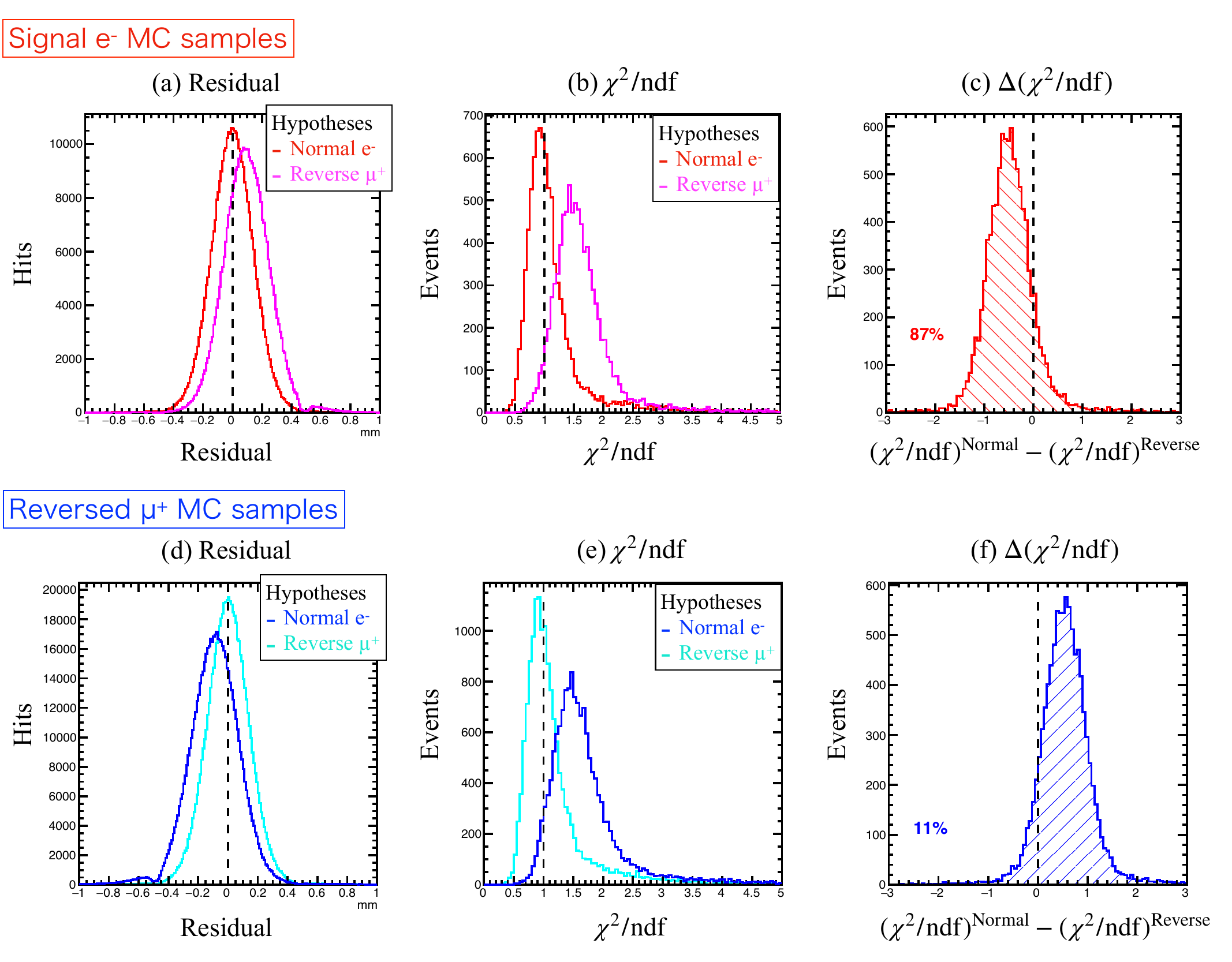}
  \caption{Track fitting results for the signal $e^-$ Monte Carlo samples (a, b, c), and the reversed $\mu^+$ samples (d, e, f). In the top plots (a, b), the red (magenta) histograms show the results with the normal $e^-$ (reverse $\mu^+$) hypothesis. In the bottom plots (d, e), the blue (cyan) histograms show the results with the normal $e^-$ (reverse $\mu^+$) hypothesis.}
  \label{fig:residual_chi2}
\end{figure}

The tracks were reconstructed using GENFIT2 as described in Sec.~\ref{sec:track_recon}. 
The track fitting was applied twice for an identical event based respectively on two hypotheses of the normal and reverse track directions. 
Figure~\ref{fig:residual_chi2}(a) shows the residual distributions for the signal electron samples assuming the normal and reverse directions. 
The residual is defined by the difference between the drift distance and the closest distance of track to the wire. 
The track fitting with normal $e^-$ hypothesis (the red histogram) gives the reasonable symmetric residual distribution; 
in contrast, the residual distribution shifts to the right for the fitting with the reverse $\mu^+$ hypothesis (the magenta histogram) due to the wrong assumption. 
Figure~\ref{fig:residual_chi2}(b) shows the reduced $\chi^2$ distributions for the signal electron samples with the two hypotheses. 
The reduced $\chi^2$ with the reverse $\mu^+$ hypothesis (the magenta histogram) becomes worse due to the wrong assumption. 
Figure~\ref{fig:residual_chi2}(c) shows the difference of the reduced $\chi^2$'s with the two hypotheses for an identical event, defined as 
$$
\Delta(\chi^2/\mathrm{ndf}) = (\chi^2/\mathrm{ndf}) ^\mathrm{Normal} - (\chi^2/\mathrm{ndf}) ^\mathrm{Reverse} . 
$$
The median peak shifting to the left can be seen. 

Figure~\ref{fig:residual_chi2}(d), (e) and (f) show the results for reversed $\mu^+$ samples based on the same analysis as Fig.~\ref{fig:residual_chi2}(a), (b) and (c), respectively. 
Note that the normal $e^-$ hypothesis is a wrong assumption in this samples, and thus the reduced $\chi^2$ difference shifts to the opposite side in Fig.~\ref{fig:residual_chi2}(f) compared with Fig.~\ref{fig:residual_chi2}(c). 

\begin{figure}[tbh] 
  \includegraphics[width=\linewidth]{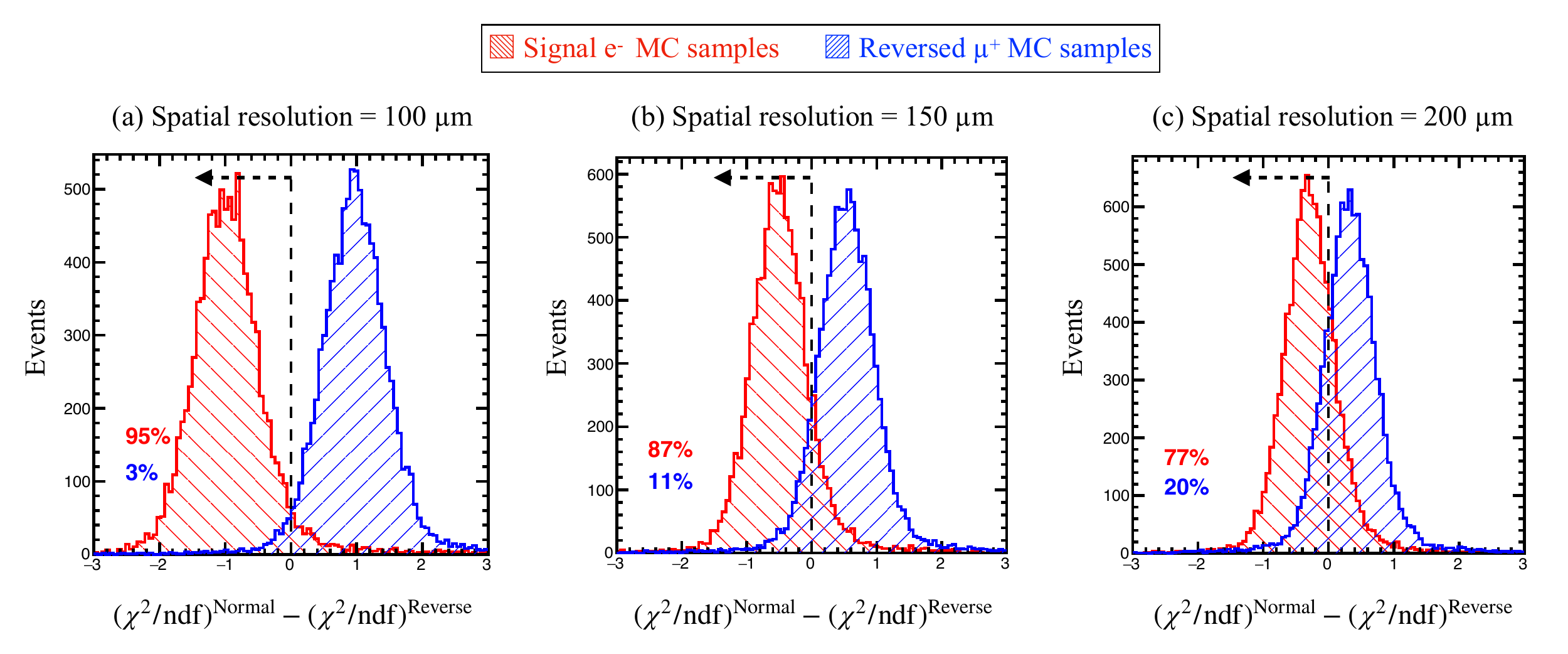}
  \caption{$\Delta(\chi^2/\mathrm{ndf})$ distributions obtained for the spatial resolution of (a) 100, (b) 150, and (c) 200~$\mu$m. The red (blue) histograms show the results with the normal $e^-$ (reversed $\mu^+$) samples. The vertical scales are arbitrary. The signal retention and background suppression rates are also displayed in red and blue, respectively, given the threshold set by $\Delta(\chi^2/\mathrm{ndf}) < 0$.}
  \label{fig:signal-vs-bg}
\end{figure}

The overlay of the $\Delta(\chi^2/\mathrm{ndf})$ distributions for the signal electron samples (Fig.~\ref{fig:residual_chi2}(c)) and the reversed muon samples (Fig.~\ref{fig:residual_chi2}(f)) is shown in Fig.~\ref{fig:signal-vs-bg}(b). 
Given a proper threshold, {\it e.g.} $\Delta(\chi^2/\mathrm{ndf}) < 0$, 87\% of the signal electrons are retained; and 89\% of the reversed $\mu^+$ are eliminated, in other words the background can be suppressed into 11\%. 

The results above are obtained by using the expected CDC spatial resolution of 150~$\mu$m. 
The dependence on the spatial resolution was also studied. 
Fig.~\ref{fig:signal-vs-bg}(a), (b) and (c) show the $\Delta(\chi^2/\mathrm{ndf})$ distributions obtained for the spatial resolution of 100, 150 and 200~$\mu$m, respectively. 
The signal retention and background suppression rates are summarized in Table~\ref{tab:res_dep}. 
The dependence on the spatial resolution is not negligible; 
therefore keeping the expected resolution of 150~$\mu$m, moreover pursuing better resoltuion, would be important. 

\begin{table}[h] 
    \centering
    \begin{tabular}{lccc}
        \hline
        Spatial resolution [$\mu$m] & 100 & 150 & 200 \\
        \hline
        Signal $e^-$ retention [\%] & 95 & 87 & 77 \\
        Background $\mu^+$ suppression [\%] &  3 &  11 & 20  \\
        \hline
    \end{tabular}
    \caption{Signal $e^-$ retention and background $\mu^+$ suppression rates in the case of the CDC spatial resolution of 100, 150 and 200~$\mu$m.}
    \label{tab:res_dep}
\end{table} 

\section{Summary and Prospects}

One of the most challenging backgrounds in the COMET Phase-I experiment originates from the sneaking cosmic $\mu^+$ that enters through an uncovered loophole of the detector solenoid magnet. 
These positive muons can scatter and produce tracks in the cylindrical drift chamber, mimicking a signal-like track at 105 MeV/$c$. 
To mitigate this issue, we implemented a track-direction discrimination method based on track-fitting quality using the GENFIT2 framework. 
The results demonstrated that this method reduces the positive muon background by an order of magnitude. 
It was also found that the spatial resolution of the tracking detector is sensitive to the reduction power. 

As a perspective, there is room to improve the reduction power. 
Utilizing the difference of energy deposit in the trigger plastic scintillators is promising. 
In addition to the combination of the track direction and the energy deposit, modern techniques using machine learning can be applicable. 
In particular, instead of the simple comparison of reduced $\chi^2$ reported in this study, each hit information with its order in the CDC track could be individually treated as an input to the machine learning approach.

\vspace{5mm}


\begin{thebibliography}{99}
\bibitem{COMET20} R. Abramishvili {\it et al.} (COMET Collaboration), COMET Phase-I technical design report, Prog. Theor. Exp. Phys. {\bf 2020}, 033C01 (2020). 
\bibitem{Moritsu22} M. Moritsu, Search for muon-to-electron conversion with the COMET experiment, Universe {\bf 8}, 196 (2022). 
\bibitem{Moritsu21} M. Moritsu, The COMET experiment: search for muon-to-electron conversion, JPS Conf. Proc. {\bf 33}, 011111 (2021). 
\bibitem{Oishi25} K. Oishi, The COMET experiment to search for $\mu$-$e$ conversion at J-PARC, JPS Conf. Proc. in this volume, O-052. 
\bibitem{Bertl06} W. Bertl {\it et al.} (SINDRUM-II Collaboration), A search for $\mu - e$ conversion in muonic gold, Eur. Phys. J. C {\bf 47}, 337 (2006). 
\bibitem{Sato24} A. Sato, H. Yoshida, M. Moritsu {\it et al.}, Design and construction of the cylindrical drift chamber for the COMET Phase-I experiment, Nucl. Instrum. Meth. A {\bf 1069}, 169926 (2024). 
\bibitem{Artikov24} A. Artikov, V. Baranov, A. Boikov, D. Chokheli, Yu.I. Davydov, V. Glagolev, A. Simonenko, Z. Tsamalaidze, I. Vasilyev, and I. Zimin, High efficiency muon registration system based on scintillator strips, Nucl. Instrum. Meth. A {\bf 1064}, 169436 (2024). 
\bibitem{Niess22} V. Niess, The PUMAS library, Computer Physics Communications {\bf 279}, 108438 (2022). 
\bibitem{Fujii24} Y. Fujii, R. Sasaki {\it et al.}, Particle identification using plastic scintillators in the COMET Phase-I experiment, Nucl. Instrum. Meth. A, {\bf 1067}, 169665 (2024). 
\bibitem{Wu21} C. Wu, T.S. Wong {\it et al.}, Test of a small prototype of the COMET cylindrical drift chamber, Nucl. Instrum. Meth. A {\bf 1015}, 165756 (2021). 
\bibitem{Moritsu19} M. Moritsu {\it et al.}, Construction and performance tests of the COMET CDC, Proceedings of Science {\bf 340} (ICHEP2018), 538 (2019). 
\bibitem{Moritsu20} M. Moritsu {\it et al.}, Commissioning of the cylindrical drift chamber for the COMET experiment, Proceedings of Science {\bf 364} (EPS-HEP2019), 128 (2020). 
\bibitem{PDG24} S. Navas {\it et al.} (Particle Data Group), The Review of Particle Physics, Phys. Rev. D {\bf 110}, 030001 (2024). 
\bibitem{Kaneko24} F. Kaneko, Y. Kuno, J. Sato, I. Sato, D. Pieters, and C. Wu, Extracting signal electron trajectories in the COMET Phase-I cylindrical drift chamber using deep learning, Prog. Theor. Exp. Phys. {\bf 2025}, 053C01 (2025). 
\bibitem{Zhang19} Y. Zhang, Y. Nakatsugawa, H.B. Li, Y. Yuan, and T.Y. Xing, Multi-turn track fitting for the COMET experiment, EPJ Web of Conf. {\bf 214}, 02005 (2019). 
\bibitem{Hoppner10} C. H\"{o}ppner, S. Neubert, B. Ketzer, and S. Paul, A novel generic framework for track fitting in complex detector systems, Nucl. Instrum. Meth. A 620, 518 (2010).
\bibitem{Bilka19} T. Bilka et al., Implementation of GENFIT2 as an experiment independent track-fitting framework, arXiv:1902.04405 (2019). 
\end{thebibliography}
\end{document}